\documentclass[11pt]{article}
\usepackage{amsfonts}
\usepackage{amssymb}
\usepackage{amsmath}
\usepackage{amsthm}
\usepackage{epsfig}

\newlength{\bredde}
\def\slash#1{\settowidth{\bredde}{$#1$}\ifmmode\,\raisebox{.15ex}{/}
\hspace*{-\bredde} #1\else$\,\raisebox{.15ex}{/}\hspace*{-\bredde} #1$\fi}
\newcommand{\be}{\begin{equation}}
\newcommand{\ee}{\end{equation}}
\newcommand{\bea}{\begin{eqnarray}}
\newcommand{\eea}{\end{eqnarray}}
\newcommand{\nn}{\nonumber}
\newcommand{\eins}{\leavevmode\hbox{\small1\kern-3.8pt\normalsize1}}
\newcommand{\e}{\mbox{e}}

\newcommand{\erfc}{\mbox{erfc}}
\newcommand{\hR}{\widehat{R}}
\newcommand{\sect}[1]{\setcounter{equation}{0}\section{#1}}

\def\Tr{{\mbox{Tr}}}

\begin{document}
\topmargin -1.4cm
\oddsidemargin -0.8cm
\evensidemargin -0.8cm
\title{\Large{{\bf
Universal microscopic correlation functions for products of
independent Ginibre matrices}}}

\vspace{1.5cm}

\author{~\\{\sc Gernot Akemann}$^{1}$ and {\sc Zdzislaw Burda}$^{2}$
\\~\\
$^1$Department of Physics,
Bielefeld University,\\
Postfach 100131,
D-33501 Bielefeld, Germany\\~\\
$^2$Marian Smoluchowski Institute of Physics, Jagellonian University,\\
Reymonta 4, 30-059 Kr\'akow, Poland
}
\date{}
\maketitle
\vfill

\begin{abstract}
We consider the product of $n$ complex non-Hermitian, independent
random matrices, each of size $N\times N$ 
with independent identically distributed Gaussian entries (Ginibre matrices).
The joint probability distribution of the complex eigenvalues of the
product matrix is found to be
given by a determinantal point process as in the case of a single
Ginibre matrix, but with a more complicated weight given by a Meijer
$G$-function depending on $n$. Using the method of orthogonal polynomials we
compute all eigenvalue density correlation functions exactly for finite $N$ and
fixed $n$. They are given by the determinant of the corresponding kernel which
we construct explicitly. In the large-$N$ limit at fixed $n$ we first determine
the microscopic correlation functions in the bulk and at the edge of
the spectrum. After unfolding they are identical to that of the
Ginibre ensemble with $n=1$ and thus universal. In contrast the
microscopic correlations we find at the origin differ for each 
$n>1$ and
generalise the known Bessel-law in the complex plane for $n=2$ to a
new hypergeometric kernel 
$\mbox{}_0F_{n-1}$.

\end{abstract}

\vfill

\thispagestyle{empty}
\newpage

\renewcommand{\thefootnote}{\arabic{footnote}}
\setcounter{footnote}{0}

\sect{Introduction}\label{intro}

The topic of random matrix theory (RMT) enjoys an increasing number of
applications in physics, mathematics and other sciences, and we refer
to \cite{handbook} for a recent compilation. This statement holds both
for problems with real eigenvalues as well as with complex
eigenvalues. In the latter case the classical ensemble was introduced
by Ginibre \cite{Ginibre} who considered complex
non-Hermitian matrices $X$ of size $N\times N$, with all matrix elements
having independent normal distributions.
However, for certain applications it is not sufficient to introduce a
single random 
matrix, e.g. when considering transfer matrices. The problem of
studying such products of random matrices is as old as RMT itself and
was introduced by 
Furstenberg and Kesten \cite{FK}.
Applications for products of random matrices range from combinatorics
\cite{PZ} over Lyapunov exponents in diffusion problems \cite{peter} to
Quantum Chromodynamics \cite{Osborn} to name a few recent works.

One problem one has to face is that the product matrix often has less
symmetry than the individual matrices. For example the product $P$ of two
Hermitian matrices is in general no longer Hermitian, and thus acquires
a complex spectrum. Therefore in the literature two types of products
have been considered, those which are Hermitised by considering
$P^\dag P$ with real spectra, and those which are studying the
complex eigenvalues of $P$ itself. In this paper we will consider the
latter problem in the simplest setting, by multiplying a fixed
number $n$ of independent Ginibre matrices: $P_n=X_1X_2\ldots X_n$.
This problem has found a renewed interest in recent years in the
mathematics and physics literature. In particular the mean or
macroscopic density of
complex eigenvalues given by the $n$-th power of the circular law was
derived independently using diagrammatic methods \cite{Burda},  
singular values \cite{goetze} 
or empirical spectral distributions \cite{Burda,Burda-rect}. 
Moreover it was shown that
the product of a wide class of non-Hermitian Wigner matrices
with independent identically distributed (iid) non-Gaussian entries has 
the same spectral distribution in the large-$N$ limit as the product of
Gaussian matrices \cite{OS}.
The same macroscopic universality was observed numerically in \cite{Burda-rect}
multiplying matrices from different symmetry classes.

The aim of the present paper is to provide more detailed information
for the product matrix, by exactly solving the problem at finite matrix size
$N$ first, using orthogonal polynomials in the complex plane.
This goal was previously achieved for $n=2$ complex non-Hermitian
\cite{Osborn}, quaternion real \cite{A05} and real asymmetric matrices
\cite{APSo}. These can be considered as non-Hermitian
generalisations of Wishart-Laguerre ensembles, see \cite{GA11} for a
review.
From \cite{Osborn}
we will borrow the idea of parameterising the $n$ matrices $X_j$ 
in order to compute the joint probability distribution function
(jpdf) of $P_n$. Our approach
enables us to identify various large-$N$ limits and to resolve the
fluctuations in 3 different regions: the origin, bulk and edge.
A new origin limit was previously found in \cite{Osborn} for $n=2$,
see also \cite{APS} for gap probabilities 
in that ensemble. 
The limiting microscopic 
density at the edge for $n=2$ was shown in \cite{KS} to coincide with
that of the Ginibre ensemble $n=1$ \cite{FH,EK}, 
and was conjectured to
universally hold for arbitrary $n$, based on numerics and including
rectangular matrices with various symmetries as well \cite{Burda-rect}.
The same universal edge scaling limit can also be found for truncations of 
unitary matrices \cite{BHJ}, sub-unitary matrices \cite{EB}, 
and for the induced Ginibre ensemble \cite{Jonit}.

We find that at the origin each $n>1$ leads to a new class of
hypergeometric kernels, whereas in the bulk and at the edge the
correlations agree for all $n$ with the Ginibre ensemble
and are thus universal.
Our findings add another facet to the property of universality of
correlation functions in the complex plane.
This question was answered for a single matrix, either 
with non-Gaussian invariant distributions in the bulk \cite{ameur},
or with non-invariant iid elements
\cite{TV} in the bulk and at the edge.

The paper is organised as follows. In the next section \ref{main}
we briefly list the main results obtained in this paper.
A detailed derivation of the jpdf is given in section \ref{jpd},
including our matrix parametrisation
and the computation of the weight function in the complex plane.
Section \ref{OP} provides the complete solution for finite-$N$ at fixed $n$ for 
all eigenvalue density correlation functions in terms of orthogonal
polynomials and their kernel.
The following sections are dealing with the large-$N$ limit, first for
the density at large matrix size and large argument in section \ref{rho_macro}.
Section \ref{micro} is devoted to the three microscopic limits
mentioned earlier, before concluding in section \ref{conc}. Several
technical details are reported in the appendices \ref{JacobianA} -
\ref{nGinibreD}.

\sect{Main results}\label{main}

Given the product $P_n$ of $n$ independent matrices $X_j$, $j=1,\ldots,n$,
each of size $N\times N$
drawn from the Ginibre ensemble with Gaussian distribution proportional to
$\exp[-\Tr X_j^\dag X_j]$,
\be
P_n\equiv X_1X_2\ldots X_n\ .
\ee
The partition function $Z_n$ can be expressed as an integral of  the
joint probability distribution 
function ${\cal P}_{jpdf}$
of the complex eigenvalues $z_j$, $j=1,\ldots,N$ of  $P_n$, given by
\be
Z_n=C_n \int \prod_{a=1}^N \left( d^2z_{a} w_n(z_a)\right)
\prod^N_{b>a} \big|z_b - z_a\big|^2
\equiv \int \prod_{a=1}^N d^2z_{a} {\cal  P}_{jpdf}(\{z\})\ ,
\ee
where $C_n$ is some known constant. The weight function  $w_n(z)$ that
depends only on the modulus
is given by the so-called Meijer $G$-function
\be
w_n(z)
= \pi^{n-1} G^{n\,0}_{0\,n}\left(\mbox{}_{\vec{0}}^{-} \bigg| \, |z|^2
\right) \ .
\ee
The corresponding kernel of polynomials orthonormal with respect to
that weight reads
\be
K_N^{(n)}(z_i,z_j)=\sqrt{w_n(z_i)w_n(z_j)}
\sum_{k=0}^{N-1} \frac{1}{(\pi k!)^n} (z_iz_j^*)^k\ .
\label{KNdef}
\ee
The $k$-point density correlation functions then easily follow to be
the determinant of that kernel,
\bea
R^{(n)}_k(z_1,\ldots,z_k)&\equiv& \frac{N!}{(N-k)!}
\frac{1}{Z_n}
\int d^2z_{k+1}\ldots   d^2z_{N} {\cal P}_{jpdf}(\{z\})
\label{Rkdef}\\
&=&\det_{1\leq i,j\leq k}\left[K_N^{(n)}(z_i,z_j)\right]\ .
\label{Rkdet}
\eea
After briefly summarising our results for both finite $n$ and $N$
which are new for $n>2$ we turn to the large-$N$ limits.

For large $N$ and large arguments $|z|\gg1$ the eigenvalue density behaves
as
\begin{equation}
R^{(n)}_1(z) =K_N^{(n)}(z,z)\approx \frac{|z|^{\frac{2}{n}-2}}{n\pi}
\frac{1}{2}\mbox{erfc}\left(\frac{\sqrt{n}(|z|^{2/n} - N)}{\sqrt{2}
  |z|^{1/n}} \right).
\label{R1large}
\end{equation}
From this the mean or macroscopic large-$N$ density can be read off:
\be
\rho^{(n)}_{macro}(w) \equiv\lim_{N\to\infty}\frac1N N^nR_1^{(n)}(z=N^{n/2}w)
= \frac{|w|^{\frac{2}{n}-2}}{n\pi}
\Theta(1-|w|) \ .
\ee
This result was previously derived in \cite{Burda,goetze,OS} 
and we have rescaled
variables such that the support is now the unit disc.
Our new results are for the microscopic limit at the origin, 
in the bulk and at the
edge of the spectrum. We first quote the latter, by zooming into the
region around the edge of the support, which is $z\approx
N^{\frac{n}{2}}$ in eq. (\ref{R1large}):
\be
\rho^{(n)}_{edge}(\xi) \equiv \lim_{N\to\infty}
nN^{n-1}R_1^{(n)}\left(z=N^{n/2}(1+\xi\sqrt{n/N})\e^{i\phi}\right)
=\frac{1}{2\pi}\erfc(\sqrt{2}\xi)\ .
\ee
This result depending only on the radial distance from the edge
is universal in the sense that it agrees for all $n$ with that of
the Ginibre ensemble \cite{FH,EK} with $n=1$ (see also \cite{Mehta}), 
and the non-Hermitian Wishart
ensemble \cite{KS} with $n=2$ derived previously.

For the bulk scaling limit we first have to unfold, in order to have a
flat density locally. In our case this can be obtained by a simple
rescaling $\xi=\sqrt{n}z^{1/n}$,
\be
\hR_1^{(n)}(\xi)=\frac1n
z^{2-\frac{2}{n}}R_1^{(n)}\left(z=(\xi/\sqrt{n})^{n}\right) \ ,
\ee
and correspondingly for higher correlation functions. Taking the
scaling limit $1\ll \xi\lesssim \sqrt{N}$ to be in the bulk 
we obtain the
following answer for the asymptotic kernel
\bea
K_{bulk}^{(n)}(\xi_i,\xi_j)&\equiv&
\lim_{N\to\infty;|\xi_{i,j}|\gg1}n^{2-n}|\xi_i\xi_j|^{n-1}K_N^{(n)}\left(
z_i=(\xi_i/\sqrt{n})^{n},z_j=(\xi_j/\sqrt{n})^{n}
\right)\nn\\
&=&\frac1\pi \left(\frac{\xi_i\xi_j^*}{|\xi_i\xi_j|}\right)^{(1-n)/2}
\exp\left[-\frac12(|\xi_i|^2+|\xi_j|^2+\xi_i\xi_j^*)\right].
\eea
This is equal to the rescaled universal kernel in the bulk of the Ginibre
ensemble
$n=1$, in which the bulk and origin limits coincide. 
The phase factor in front of the exponential 
is irrelevant as it drops out
after taking the determinant in eq. (\ref{Rkdet}).

Finally let us emphasise that there exists a third microscopic
large-$N$ limit
at the origin which {\it differs} for every
$n>1$. Its kernel is simply given by extending the sum in
eq. (\ref{KNdef}) to infinity, leading to a hypergeometric function:
\be
K_{origin}^{(n)}(z_i,z_j)\equiv\lim_{N\to\infty}K_N^{(n)}(z_i,z_j)
=\sqrt{w_n(z_i)w_n(z_j)}\
\mbox{}_0F_{n-1}(-;\vec{1};z_iz_j^*)\,\pi^{-n}\ , n>1\ .
\ee
Here we have $|z_i|={\cal O}(1)$.
For the Ginibre ensemble $n=1$ the limiting kernel is given by that of
the bulk limit
above, whereas for $n=2$ one has a $K$-Bessel function for the weight
times an $I$-Bessel function from the sum in the kernel \cite{Osborn}, 
which was also investigated in \cite{APS} including its Fredholm determinant.

\sect{Derivation of the joint eigenvalue distribution}\label{jpd}

\subsection{Outline of the method}
We are interested in statistical properties of eigenvalues of the product
\begin{equation}
P_n = X_1 X_2 \ldots X_n
\label{P}
\end{equation}
of $n$ independent Ginibre matrices of dimensions $N\times N$.
The partition function  of independent matrices factorises into
a product of independent partition functions for each matrix.
We however write it in a compact way
\begin{equation}
Z_n= \prod_{j=1}^n \int |DX_j| \ \e^{- {\rm Tr} X^\dagger_j X_j} =
\int |DX|\ \e^{- {\rm Tr} \sum_{j=1}^n X^{\dagger}_j X_j} \ .
\label{Z}
\end{equation}
because we are interested in the eigenvalues of $P_n$.
In this notation $DX$ is the Euclidean volume form (external
product of all independent one-forms) and $|DX|$ is
the corresponding unoriented volume element:
\begin{equation}
\begin{split}
|DX| = \prod_{j=1}^n |DX_j| & = \prod_{j=1}^n \prod_{a,b=1}^N
\left(\left(d {\rm Re} X_j\right)_{ab} \left(d {\rm Im}
X_j\right)_{ab}/2\pi\right) \\
& = \prod_{j=1}^n \prod_{a,b=1}^N
\left(\left(d  X_j\right)_{ab} \left(d X_j\right)^*_{ab}/4\pi\right)\ .
\end{split}
\end{equation}
The star denotes the complex conjugate.
The index $j$ runs over the set of matrices $j=1,2,\ldots, n$ and
$a$ and $b$ over rows and columns of the matrix $a =1,\ldots,N$ and
$b=1,\ldots,N$. 

Our first goal is to rewrite the integrand of (\ref{Z}) in new variables
that can be divided into two classes: the first one comprising variables
that are directly related to the eigenvalues of $P_n$ and  the second one
comprising variables independent of the eigenvalues of $P_n$. Having done
that we can try to integrate out the variables of the second class to obtain
an integrand depending only on eigenvalues of $P_n$. We obtain in this way
the joint probability distribution function ${\cal P}_{jpdf}$
for eigenvalues of $P_n$. We basically
follow the Dyson procedure but introduce some pertinent modifications
that enable us to handle the case of eigenvalues of the product matrix
$P_n$ in eq. (\ref{P}).

It is convenient to consider a cyclic block matrix $B$, following \cite{Burda}:
\begin{equation}
B=\left(
\begin{array}{cccccc}
0     & X_1 & 0     & 0     & \ldots & 0 \\
0     & 0     & X_2 & 0     & \ldots & 0 \\
0     & 0     & 0     & X_3 & \ldots & 0 \\
       &        &       &        & \ldots &    \\
0     & 0     & 0     & 0     & \ldots & X_{n-1} \\
X_n & 0     & 0     & 0     & \ldots & 0 \\
\end{array}
\right).
\end{equation}
Each block $X_j$ in $B$ is of dimensions $N\times N$,
so the whole matrix has dimensions $nN\times nN$. One can easily
see that the $n$-th power 
$B^n$
has identical eigenvalues as $P_n$, but  each being
$n$ times degenerate. Indeed, $B^n$ is a block diagonal matrix.
The first diagonal block is $X_1X_2\ldots X_n$,
the second one is $X_2X_3\ldots X_nX_1$, and all others are products
of consecutive cyclic permutations of $X_1$, $X_2$, $\ldots$, $X_n$.
Each diagonal block of $B^n$ has the same eigenvalues as $P_n$,
so therefore each eigenvalue of $P_n$ appears in $B^n$ in $n$ copies.

\subsection{Change of variables}
In this section we change variables, as outlined above, and find the
corresponding Jacobian. Without loss of generality we illustrate our
method for $n=3$. We use the following parametrisation of the block matrix:
\begin{equation}
\begin{split}
B & = \left(\begin{array}{ccc}
0 & X_1 & 0 \\
0 & 0 & X_2 \\
X_3 & 0 & 0
\end{array}\right)
\\ & = \left(\begin{array}{ccc}
U_1 & 0 & 0 \\
0 & U_2 & 0 \\
0 & 0 & U_3
\end{array}\right)
\left(\begin{array}{ccc}
0 & \Lambda_1 + T_1 & 0 \\
0 & 0 & \Lambda_2 + T_2 \\
\Lambda_3 + T_3 & 0 & 0 \\
\end{array}\right)
 \left(\begin{array}{ccc}
U^{-1}_1 & 0 & 0 \\
0 & U^{-1}_2 & 0 \\
0 & 0 & U^{-1}_3
\end{array}\right).
\end{split}
\label{BLTU}
\end{equation}
This is a generalisation of Osborn's idea who considered the chiral
block matrix for $n=2$ \cite{Osborn}. It is also known as $QZ$ or
generalised Schur decomposition, see e.g. \cite{GVL} section 7.7.
The diagonal block unitary matrix $U$ on the left hand side has three
diagonal unitary blocks $U_j$ of dimensions
$N\times N$ restricted to the right coset $U(N)/U(1)^N$.
Each block $\Lambda_j$ is a $N\times N$ diagonal matrix
$\Lambda_j = \mbox{diag}(x_{j1},\ldots,x_{jN})$. Each block $T_j$ is
an $N\times N$ strictly upper triangular matrix. One should note that
the eigenvalues of $B^3$ (and thus also of $P_3$)
are equal to $x_{1a}x_{2a}x_{3a}$ for $a=1,\ldots,N$.

In the original representation, the block matrix $B$ is parametrised
by the elements of the matrices $X_j$. Each matrix $X_j$ has $N^2$ elements,
and each element is
a complex variable (two real degrees of freedom). So altogether there
are $6N^2$ real degrees of freedom. This number matches the
number of  degrees of freedom of the parametrisation on the right-hand-side:
each $U_j$ has $N(N-1)$ real parameters, each $\Lambda_j$ has $2N$
and each $T_j$ has $N(N-1)$. It makes $2N^2$ times $3$, that is $6N^2$
real degrees of freedom as in the original parametrisation. In Appendix A
we calculate the Jacobian for the change of variables (\ref{BLTU})
and integrate out the $T_j$'s and $U_j$'s. The corresponding integrals
factorise and the constants can be read off from the
known Ginibre ensemble. We are left with an integral
\begin{equation}
Z_{3}=C_{3} \int \prod_{j=1}^n\prod_{a=1}^N d^2x_{ja} \prod_{b>a}^N
\big|x_{1b} x_{2b} x_{3b} - x_{1a} x_{2a} x_{3a}\big|^2\
\e^{- \sum_{j=1}^n\sum_{a=1}^N
  |x_{ja}|^2},\ \ C_3=\frac{1}{(N!\,G[N+1]\pi^N)^3}\ ,
\end{equation}
that depends on eigenvalues of the three matrices in the product.
Here $G[N+1]=\prod_{k=0}^{N-1}k!$ is the Barnes $G$-function.
As discussed 
the $z_a=x_{1a} x_{2a} x_{3a}$, $a=1,\ldots,N$
correspond to the complex eigenvalues of the matrix $P_n$.
So we can now rewrite the last formula by integrating out the $x_{ja}$'s,
and 
by expressing the result in terms of the $z_a$'s. This gives
the standard form of the partition function
\begin{equation}
Z_3=C_3 \int \prod_{a=1}^N \left( d^2z_{a} w_3(z_a)\right)  \prod_{b>a}^N
\big|z_b - z_a\big|^2\ ,
\end{equation}
however with a non-standard weight function:
\begin{equation}
w_3(z) = \int d^2x_{1} d^2x_{2} d^2x_{3}\ \delta^{(2)}(z - x_1x_2 x_3)
\ \e^{-\left(|x_1|^2 + |x_2|^2 + |x_3|^2\right)} \ .
\label{w3}
\end{equation}

\subsection{The weight function}

It is easy to make the generalisation of the previous two subsections to
the product (\ref{P})
of any number of matrices $n$.
Apart from the constants the only difference is that the weight
function $w_3(z)$ has to be replaced by $w_n(z)$:
\begin{equation}
Z_n=C_n \int \prod_{a=1}^N \left( d^2z_{a} w_n(z_a)\right)  \prod^N_{b<a}
\big|z_b - z_a\big|^2 ,\ \ C_n=\frac{1}{(N!\,G[N+1]\pi^N)^n}=(C_1)^n\ ,
\label{Zn}
\end{equation}
that is given by:
\begin{equation}
w_n(z) = \int \prod_{j=1}^n d^2 x_j\
\delta^{(2)}\left(z - \prod_{j=1}^n x_j\right)
\e^{- \sum_{j=1}^n |x_j|^2} \ .
\label{wn}
\end{equation}
If one introduced a probabilistic normalisation to the weight function,
the last equation could be interpreted in terms of the probability
density function of a random variable obtained being a product of
$n$ independent Gaussian complex random variables - very
close to the spirit of our original problem. Of course things are
more complicated since in addition to the weights we have also the
repulsion coming from the Vandermonde determinants. The second
observation is that the moments of the weight function factorise
into $n$ independent Gaussian integrals
\begin{equation}
\int d^2z |z|^{2k} 
w_n(z) = \prod_{j=1}^n \int d^2x_j |x_j|^{2k} 
\e^{-|x_j|^2} =  \left( \int d^2x |x|^{2k} 
\e^{-|x|^2} \right)^n=(\pi k!)^n\ ,
\label{moments_kl}
\end{equation}
since integrating out the delta constraint in (\ref{wn}) makes the
$x_j$-integrals mutually independent.

One can find an integral representation of the weight function (\ref{wn})
by integrating out one variable, say $x_{n}=z/(x_1\ldots x_{n-1})$, 
and getting rid of the delta function constraint.
Equivalently it can be done using the formula (\ref{moments_kl})
for $k=0$ in polar coordinates:
\be
r_{n}\to \frac{z}{r_{1}\cdots r_{n-1}}=\frac{r\exp[i\phi]}{r_1\cdots
  r_{n-1}}\ ,\ \ n>1\ ,
\ee
where $r_j=|x_j|$, $r=|z|$. This leads to the following transformation:
\bea
&&\int d^2z\, w_n(z)
=\int d^2x_1\ldots d^2x_n \exp\left[ -\sum_{j=1}^n |x_j|^2\right]\nn\\
&&= \int_{0}^{2\pi}d\phi \int_{0}^\infty dr\,r (2\pi)^{n-1}
\int_{0}^\infty dr_1r_1\ldots dr_{n-1}r_{n-1}\frac{1}{(r_1\cdots r_{n-1})^2}
\exp\left[ -\frac{r^2}{(r_1\cdots r_{n-1})^2} -\sum_{j=1}^{n-1}
  r_j^2\right],\nn\\
\eea
with
\be
w_n(z)=(2\pi)^{n-1}\int_{0}^\infty \frac{dr_1\ldots dr_{n-1}}{r_1\cdots r_{n-1}}
\exp\left[ -\frac{|z|^2}{(r_1\cdots r_{n-1})^2} -\sum_{j=1}^{n-1}
  r_j^2\right]\ , \ \ n>1\ .
\label{wn-def}
\ee
Obviously the weight given by an $(n-1)$-fold integral only depends on
the modulus $|z|$. It immediately follows from (\ref{wn-def}) that
\be
w_{n+1}(z)=2\pi\int_{0}^\infty \frac{dr}{r}
w_{n}(z/r)\exp\left[ - r^2\right]\ , \ \ n>0\ ,
\label{wn-rec}
\ee
with $w_1(z) = \e^{-|z|^2}$ which is the Ginibre weight.
This recursion relation can be solved for $w_{n+1}(z)$ by
using the Mellin transform in the variable $R=|z|^2$. Defining
$\Omega_{n}(R) \equiv w_{n}(\sqrt{R})$ we calculate
the corresponding transform:
\be
M_{n}(s) = \int_0^\infty dR\  R^{s-1} \Omega_{n}(R) \ .
\ee
The recursion relation (\ref{wn-rec}) takes the form
\be
\Omega_{n+1}(R)=\pi\int_{0}^\infty \frac{d\rho}{\rho}
\Omega_{n}(R/\rho) \e^{- \rho} \ ,
\ee
and its Mellin transform factorises
\be
M_{n+1}(s) = \pi M_n(s) \Gamma(s)\ ,
\ee
where $\Gamma(s) = \int_0^\infty d\rho \rho^{s-1} \e^{-\rho}$.
Together with the initial condition
\be
M_1(s)= \int_0^\infty dR\  R^{s-1}\exp[-R]=\Gamma(s)
\ee
we thus have
\be
M_{n}(s) = \pi^{n-1} \Gamma^n(s) \ .
\ee
We can now write the inverse transform
\be
w_n(z) = \Omega_n(|z|^2) =
\pi^{n-1}
\frac{1}{2\pi i} \int_{c-i\infty}^{c+i\infty} \Gamma^n(s) |z|^{-2s} ds
= \pi^{n-1} G^{n\,0}_{0\,n}\left(\mbox{}_{\vec{0}}^{-} \bigg| \, |z|^2 \right),
\label{wn-GMn}
\ee
where the symbol $G^{n\,m}_{p\,q}(\ldots |z)$ denotes Meijer's
$G$-function \cite{Grad},
and $\vec{0}$ is
a string of $n$-zeros. Only the subclass $G^{n\,0}_{0\,n}$ of these
functions, corresponding to the inverse Mellin transform of $\Gamma^n(s)$,
appears in this context. Note that our weight functions differ
from the Meijer $G$-functions  
$G^{q\,0}_{q\,q}$ appearing in the distribution of Fuss-Catalan
numbers \cite{PZ}. 

For $n=1$ this result 
reproduces by construction
the weight for Ginibre matrices (or by inverse Mellin
transform \cite{Grad})
\be
w_1(z)=\exp\left[ -|z|^2\right]=G^{1\,0}_{0\,1}\left(\mbox{}_{0}^{-}
\bigg| \ |z|^2 \right) \ . 
\label{w1-Gin}
\ee
For $n=2$ the integral eq. (\ref{wn-def})
can be performed analytically, leading to a
modified Bessel-function \cite{Osborn}:
\be
w_2(z)\ = 2 \pi K_0(2|z|)=  \pi\ 
G^{2\,0}_{0\,2}\left(\mbox{}_{0,0}^{-} \bigg| \ |z|^2 \right)\ .
\label{w2-Osb}
\ee
Alternatively one can use the connection of Meijer $G$-functions to
special functions in \cite{Grad}.
For $n=3$ we either keep a single integral over the previous
$K$-Bessel function, or we write the result explicitly in terms of the
Meijer $G$-function:
\be
w_3(|z|)\ =\ 4\pi^2 \int_{0}^\infty
\frac{dr}{r}K_0\left(2|z|/r\right)\exp[-r^2]\ =\ \pi^2\ 
 G^{3\,0}_{0\,3}\left(\mbox{}_{0,0,0}^{-} \bigg| \ |z|^2 \right)
\ .
\label{w3-MG}
\ee
For an alternative representation in terms of
hypergeometric functions and series of polygamma special functions we refer to 
\cite{Klauder} section V case f). 


\sect{The orthogonal polynomial approach}\label{OP}

Since the partition function $Z_n$ (\ref{Zn}) expressed in terms of
the eigenvalues of the matrix $P_n$ has the standard form of the product
of the weights times
the absolute 
value square of the Vandermonde determinant we can
use the standard technique of orthogonal polynomials in the complex plane
to determine the $k$-point correlation functions
\cite{Mehta}. For recent reviews on non-Hermitian
random matrix theory and its corresponding polynomials we refer to
\cite{BHJ,GA11}.
We refer the interested reader
to these references for details.

\subsection{Orthonormal polynomials and kernel}

From the fact that our weight $w_n(z)$ is angle-independent it
immediately follows that the corresponding orthogonal polynomials
are monic $p_k(z)=z^k$. Indeed, the integral
\be
\int d^2z\ w_n(z) z^k z^{*\, l} = \int_0^{2\pi} d\phi\ \e^{i\phi(k-l)}
\int_0^\infty dr\,r w(r) \equiv \delta_{kl} h^{(n)}_k
\ee
gives zero for $k\ne l$. Using eq. (\ref{moments_kl}) we immediately find that
the squared norms are
\be
h_k^{(n)}=(\pi k!)^n\ ,
\label{norms}
\ee
so that the corresponding kernel of {\it orthonormal} polynomials reads
\be
K^{(n)}_N(z_i,z_j)=\sqrt{w(z_i)w(z_j)}
\sum_{k=0}^{N-1} \frac{1}{(\pi k!)^n} (z_iz_j^*)^k \ .
\label{Ker-result}
\ee
In view of the results later in section \ref{origin} we call this a
truncated hypergeometric kernel, just as in the Ginibre case $n=1$
this is a truncated exponential.

Following \cite{Mehta}
this kernel determines the $k$-point correlation functions:
\be
R^{(n)}_k(z_1,\ldots,z_k)= \det_{1\leq i,j,\leq k}
\left[K^{(n)}_N(z_i,z_j)\right].
\label{Rk-result}
\ee
In particular, the one-point correlation function
or spectral density is $R_1^{(n)}(z) = K_N^{(n)}(z,z)$,
the two-point correlation function is $R_2^{(n)}(z,u) = K_N^{(n)}(z,z)
K_N^{(n)}(u,u)-K_N^{(n)}(z,u) K_N^{(n)}(u,z)$ etc.
We are now going to discuss these in the large-$N$ limit in the next section.

\sect{Eigenvalue density at large-$N$} \label{rho_macro}

In this section we study the eigenvalue density for large-$N$
(and large argument). The purpose of that is two-fold, as we will not
only find the mean or macroscopic density after a suitable rescaling.
The expression for large but finite-$N$ that we will derive below
enables us to identify the different regions in the complex plane
where we can take different microscopic limits that describe the
fluctuations in that region.

We begin with the definition of the spectral density or
one-point correlation function in terms of the kernel:
\begin{equation}
R^{(n)}_1(z)=K^{(n)}_N(z,z)= w_n(z) \, \sum_{k=0}^{N-1}
\frac{|z|^{2k}}{(\pi k!)^n} 
\equiv w_n(z) \, T_n\left(|z|^2,N\right) .
\label{TNdef}
\end{equation}
We note that according to its definition eq. (\ref{Rkdef})
it is normalised to the number of eigenvalues,
$\int d^2 z R^{(n)}_1(z)=N$, rather than to unity.

Now we are going to determine the behaviour of $R^{(n)}_1(z)$ for
large $N$ and for large $|z|$. 
The asymptotic large-$|z|$ behaviour of the weight $w_n(z)$ can be
taken from \cite{Fields}. It  is rederived here
for completeness in  Appendix \ref{weightB} (see equation (\ref{asympt})).
The large-$N$ behaviour of the truncated sum $T_n(x,N)$
for $x=|z|^2$ of order $N^n$ was already studied in \cite{BJK}, and
we give some details of the derivation in Appendix \ref{kernelC}
(see equation (\ref{T-asympt})). Putting together (\ref{asympt}) and
(\ref{T-asympt}) we obtain
the following behaviour for large-$N$ and for $|z|\lesssim N^{n/2}$:
\begin{equation}
R^{(n)}_1(z) \approx \frac{|z|^{\frac{2}{n}-2}}{n\pi}
\frac{1}{2}\mbox{erfc}\left(\frac{\sqrt{n}(|z|^{2/n} - N)}{\sqrt{2}
  |z|^{1/n}} \right).
\label{R1largeN}
\end{equation}
It is convenient to recast this result into a rescaled density with
compact support that is normalised to unity.
Using the rescaled
variable $w = z N^{-n/2}$ we define the following density, for which
the radius of the eigenvalue support approaches one for
$N\rightarrow \infty$:
\bea
\rho_n(w) &\equiv& \lim_{N\gg1} N^{n-1} R^{(n)}_1(N^{n/2} w)
= \frac{|w|^{\frac{2}{n}-2}}{n\pi} \frac{1}{2}\mbox{erfc}\left(
\frac{\sqrt{nN}(|w|^{2/n} - 1)}{\sqrt{2} |w|^{1/n}} \right) \nn\\
&=& \frac{|w|^{\frac{2}{n}-2}}{n\pi} \frac{1}{2}\mbox{erfc}
\left(\sqrt{\frac{2N}{n}}\left(|w| - 1\right)\right) \ .
\label{fs-n}
\eea
The factor $N^n$ in the first step
comes from the rescaling of the volume element
$d^2z = N^n d^2 w$, and the additional factor $N^{-1}$
from the normalisation to unity.
In the second step we have simplified $\rho_n(w)$ by
Taylor expanding the argument 
around unity, $|w| = (2/n) (|w|-1) + O((|w|-1)^2)$. This is because 
the complementary error function changes only in a narrow
strip around the unit circle $|w|=1$, of a width
proportional to $1/\sqrt{N}$.
So far we have derived a finite size form of the eigenvalue distribution of the
product of $n$ independent Ginibre matrices. This form was known for
$n=1$ \cite{FH,EK} and $n=2$ \cite{KS}.
It was conjectured to hold for any $n$
and tested numerically in
\cite{Burda-rect} except that the dependence of the coefficient inside the
erfc-function on $n$ was unknown.
We have filled this gap
here by deriving eq. (\ref{fs-n}) for any $n$.
We see that the width of the crossover region around the edge $|w|=1$
is proportional to the square root of the number of multiplied
matrices $\sqrt{n}$.

It is instructive to compare this result with the
limiting density for large-$N$ of the $n$-th power of a single Ginibre
matrix. 
This density is derived in Appendix \ref{nGinibreD} and is given by
exactly the same distribution, however
with a different dependence on $n$:
\be
\tilde{\rho}_{n}(w) =  \frac{|w|^{\frac{2}{n}-2}}{n\pi}
\frac{1}{2}\mbox{erfc}\left(\frac{\sqrt{2N}}{n}\left(|w| -
1\right)\right).
\ee
Here, the width of the crossover region
is proportional to $n$ and not to $\sqrt{n}$. In a sense, the finite
size corrections are stronger for the $n$-th power than for the
product of $n$ independent Ginibre matrices.

Actually the same seems to be true for a broader class of matrices called
isotropic \cite{BNS}. Isotropic matrices are known to have the property
that the product of 
iid matrices has
the same limiting distribution as the corresponding power of a single
one, as was discussed in \cite{BNS}. For this class the finite size
formula is not known but numerically one also observes that the
product of independent matrices converges faster to the limiting
distribution than does the corresponding power \cite{BNS}.

We close this section by rederiving the mean or macroscopic density in
the limit $N\to\infty$. It is obtained by taking the limit $N\rightarrow \infty$  of (\ref{fs-n}): 
\be
\rho_{macro}^{(n)}(w)\equiv\lim_{N\to\infty}\rho_n(w)
= \frac{|w|^{\frac{2}{n}-2}}{n\pi} \Theta(1-|w|) \ ,
\label{limiting-n}
\ee
where $\Theta(x)$ is the Heaviside step function. This result was
derived in \cite{Burda} using planar diagrams and rigorously in
\cite{goetze,OS}. 
We also refer to \cite{goetze} for an extensive list of references
regarding the mathematical literature on products of random matrices.

\sect{Microscopic large-$N$ limits and universality}\label{micro}

\subsection{Universal edge limit}

In this subsection we analyse the fluctuations of the eigenvalues around
the edge of the spectrum. We will only consider the density itself,
higher order correlation functions can be dealt with analogously.
Our starting point is the density at large-$N$ and large-$|z|$,
eq. (\ref{R1largeN}). As was already mentioned in the previous section
the complementary error-function rapidly vanishes for  $|z|>N^{n/2}$,
on a strip of width $1/\sqrt{N}$. We therefore introduce a new scaling
variable $\xi$:
\be
z\ =\ N^{n/2}\left(1+\xi\sqrt{n/N}\right)\e^{i\phi}\ .
\ee
The additional rescaling with $\sqrt{n}$ is made in order to normalise properly.
We can therefore define the microscopic density at the edge by
\be
\lim_{N\to\infty}
\rho^{(n)}_{edge}(\xi) \equiv \lim_{N\to\infty}
nN^{n-1}R_1^{(n)}\left(z=N^{n/2}(1+\xi\sqrt{n/N})\e^{i\phi}\right)
=\frac{1}{2\pi}\erfc(\sqrt{2}\xi)\ ,
\ee
where for each real dimension the density $R_1$ has to be rescaled by
$\sqrt{n}N^{n/2}$. The resulting expression agrees with the $n=1$
result for the Ginibre ensemble derived in \cite{FH,EK} (see also
\cite{Mehta}). For $n=2$ the same result was previously derived in
\cite{KS}. We have shown here that the result holds true for any fixed
$n$ and is thus universal. This behaviour was already observed
numerically in \cite{Burda-rect}.

\subsection{Universal bulk limit}

In this subsection 
we discuss the behaviour of the $k$-point correlation functions
$R^{(n)}_k(z_1,\ldots,z_k)$ defined in (\ref{Rk-result}), in the limit
$N\rightarrow \infty$ for the following scale separation $1 \ll |z_k|
\lesssim N^{n/2}$. By comparing to eq. (\ref{R1largeN}) this separation
will guarantee that we are neither close to the origin to be discussed
later, nor to the edge limit discussed above.

In principle the microscopic bulk limit is obtained in three steps.
We would first have to rescale $z=N^{n/2}w$ in order to get a compact
support. In the second step we would zoom into the fluctuations which
are in this case of the order $1/N^{n/2}$, $w=v/N^{n/2}$. In the
last step we would unfold the variables in order to get a locally flat
density, with a mean level spacing of unity.
Because of the macroscopic density being given by
\be
\rho^{(n)}_{macro}(w)
= \frac{|w|^{\frac{2}{n}-2}}{n\pi}
\Theta(1-|w|) \ ,
\ee
the unfolding would read $\xi=\sqrt{n}v^{1/n}$. Because the 
rescalings
from the first two steps compensate each other here, we can do all
steps in one. Our microscopic scaling limit is thus given by
the change of variables
\be
1\ll|\xi = \sqrt{n} z^{1/n}|\lesssim \sqrt{N}\ ,
\label{xiz}
\ee
which will become $N$-independent in the large-$N$ limit.
Note that in the variable $ z^{1/n}$ the fluctuations are again of
order $1/\sqrt{N}$ as in the Ginibre ensemble.
The change of variables for the $k$-point densities can be best seen
by considering the two-dimensional differential
\begin{equation}
d^2\xi = n^{-1} |z|^{\frac{2}{n}-2} d^2z \ .
\end{equation}
We thus obtain for the microscopic density
\be
\hR_1^{(n)}(\xi)\equiv n|z|^{2-\frac{2}{n}}
R_1^{(n)}\left(z=\left(\xi/\sqrt{n}\right)^{n}\right)
=n|z|^{2-\frac{2}{n}}\frac{1}{n\pi}
|z|^{\frac{2}{n}-2}=\frac{1}{\pi}\ ,
\ee
which is constant on a disk of radius $\sqrt{N}$ and zero otherwise.
So indeed the eigenvalues are uniformly distributed in the $\xi$-plane
for large $|\xi|$ inside the support.
More generally, the microscopic $k$-point correlation functions are
defined as
\begin{equation}
\widehat{R}_k^{(n)}(\xi_1,\ldots,\xi_k)  \equiv n^{k} |z_1\ldots
z_k|^{2-2/n} R_k^{(n)}(z_1,\ldots,z_k)\ ,
\end{equation}
where $z_j=\left(\xi_j/\sqrt{n}\right)^{n}$ on the right hand side,
from eq. (\ref{xiz}).
All $k$-point correlation functions  $\widehat{R}_k^{(n)}(\xi_1,\ldots,\xi_k)$
can be expressed in terms of a redefined kernel
\be
\widehat{R}^{(n)}_k(\xi_1,\ldots,\xi_k)= \det_{1\leq i,j,\leq
  k}\left[\widehat{K}^{(n)}_N(\xi_i,\xi_j)\right]\ ,
\label{rh_kh}
\ee
that is given by
\be
\widehat{K}^{(n)}_N(\xi_i,\xi_j)= n|z_iz_j|^{1-1/n}K^{(n)}_N(z_i,z_j)
=n^{2-n} |\xi_i\xi_j|^{n-1}
K^{(n)}_N\left(\left(\xi_i/\sqrt{n}\right)^{n},
\left(\xi_j/\sqrt{n}\right)^{n}  \
\right) \ .
\ee
In the limit $N\rightarrow \infty$ and $1\ll |\xi_j| \lesssim \sqrt{N}$
the limiting kernel in the bulk is given by
\be
\widehat{K}^{(n)}_{bulk}(\xi_i,\xi_j)\equiv \lim_{N\to\infty;|\xi_{i,j}|\gg1}
n^{2-n} |\xi_i\xi_j|^{n-1}
\sqrt{w_n\left(\left(\xi_i/\sqrt{n}\right)^{n}\right)
  w_n\left(\left(\xi_j/\sqrt{n}\right)^{n}\right)} \;
T_n\left(\left(\xi_i \xi_j^*/n\right)^n\right)\ ,
\ee
where we have for asymptotically large argument (\ref{asympt})
\be
\lim_{|\xi|\gg1}w_n\left(\left(\xi/\sqrt{n}\right)^{n}\right)
= \frac{1}{\sqrt{n}}
\left(2\pi^3 n \right)^{(n-1)/2} |\xi|^{1-n} \e^{-|\xi|^2}\ ,
\ee
and for asymptotically large argument and large-$N$ (\ref{T-asympt})
\be
\lim_{N\to\infty;|\xi|\gg1}T_n\left(\left(\xi_i \xi_j^*/n\right)^n\right) =
 \frac{1}{\pi\sqrt{n}} \left(2\pi^3\right/n)^{(1-n)/2}
\left(\xi_i \xi_j^*\right)^{(1-n)/2} \exp[\xi_i \xi_j^*]  \ .
\ee
Putting all together we obtain
\be
\widehat{K}^{(n)}_{bulk}(\xi_i,\xi_j)= \frac{1}{\pi}
\left(\frac{\xi_i\xi_j^*}{|\xi_i\xi_j|}\right)^{(1-n)/2}
\exp\left[-\frac12(|\xi_i|^2+|\xi_j|^2+\xi_i\xi_j^*)\right]
\ .
\label{micro-Kn}
\ee
We see that the kernel is Hermitian
$\widehat{K}^{(n)}(\xi_i,\xi_j)^*=\widehat{K}^{(n)}(\xi_j,\xi_i)$.
It is equal to the kernel of the Ginibre ensemble  multiplied by powers of
the phase factors $\e^{i\phi_j} = \xi_j/|\xi_j|$.
The kernel (\ref{micro-Kn}) is unitarily equivalent to the kernel of
the Ginibre ensemble. When one calculates the correlation functions
$\widehat{R}^{(n)}_k(\xi_1,\ldots,\xi_k)$ (\ref{rh_kh}) all phase factors
cancel 
and one obtains exactly the same $k$-point correlation functions as for
the Ginibre ensemble with $n=1$. Therefore all correlation functions
of the product matrix $P_n$
in the bulk limit are universal.
In particular the two-point correlation function is given by
\be
\widehat{R}^{(n)}_{bulk}(\xi_1,\xi_2) = \frac{1}{\pi^2} \left(1 -
\e^{-|\xi_1-\xi_2|^2}\right),
\ee
which describes
correlations of eigenvalues at distances $|\xi_1-\xi_2|$ of order unity.

\subsection{Microscopic origin limit: n classes}
\label{origin}

In this subsection we investigate the remaining region where a
non-trivial microscopic limit can be obtained, that is the vicinity of
the origin. While the microscopic bulk and edge limits yield universal
results for arbitrary fixed $n$, at the origin we obtain a different kernel
for each $n$. For $n=1$ it coincides with the bulk limit whereas for
$n=2$ we obtain a kernel containing Bessel-$K$ and Bessel-$I$ functions
which was known based on the work \cite{Osborn}, see also \cite{APS}
for more details.

If we look back at the previous two limits the edge correlations were
found for $|z|\approx N^{n/2}$ or $|w|\approx1$ in rescaled variables
$z=N^{n/2}w$, with fluctuations of order $1/\sqrt{N}$. The bulk limit
was obtained by considering $1\ll|z|\lesssim N^{n/2}$ or $0<|w|<1$,
with fluctuations again of order  $1/\sqrt{N}$ in terms of the
rescaled variable $\xi = \sqrt{n} z^{1/n}$.
Here we will take the limit $N\to\infty$ keeping $|z|={\cal O}(1)$, or
$|w|\approx N^{-n/2}$. Because of that the scaling is very simple in $z$: the
weight $w_n(z)$ which is $N$-independent 
remains unchanged, and 
the sum inside the kernel
eq. (\ref{TNdef}) simply has to be extended to infinity:
\be
K_{origin}^{(n)}(z_i,z_j)\equiv\lim_{N\to\infty}K_N^{(n)}(z_i,z_j)
=\sqrt{w_n(z_i)w_n(z_j)}\
\mbox{}_0F_{n-1}(-;\vec{1};z_iz_j^*)\,\pi^{-n}\ , n>1\ .
\label{Korigin}
\ee
Here we have used 
that the infinite sum eq. (\ref{TNinfinity}) is
related to a generalised hypergeometric function,
\be
T_n(x) = \sum_{k=0}^{\infty} \frac{x^k}{(\pi k!)^n} =\frac{1}{\pi^n}
\ \mbox{}_0F_{n-1}(-;\vec{1};x)\ ,
\ee
for  $n>1$. The vector $\vec{1}$ contains $n-1$ elements of unity.
For $n=1$ we simply have
\be
T_1(x) = \sum_{k=0}^{\infty} \frac{x^k}{\pi k!} =\frac1\pi \exp[x]\ .
\ee
It cancels the exponent $\exp[-|z|^2]$ coming from the weight and thus
leads to a constant density. 

For $n=2$ the result can also be expressed in terms of elementary
functions, due to
\be
T_2(x) = \sum_{k=0}^{\infty} \frac{x^k}{(\pi k!)^2}
=\frac{1}{\pi^2}I_0(2\sqrt{x})\ .
\ee
This leads to the known kernel (see e.g. \cite{APS})
\be
K_{origin}^{(n=2)}(z_i,z_j)=\frac2\pi\sqrt{K_0(2|z_i|)K_0(2|z_j|)}\
I_0(2|z_iz_j|^{\frac12})\ ,
\ee
in terms of Bessel functions, with the density
$\rho_{origin}^{(n=2)}(z)=\frac2\pi K_0(2|z|)I_0(2|z|)$.
For $n\geq3$ our kernel eq. (\ref{Korigin}) seems to be
new. Although other random matrix ensembles
with kernels containing hypergeometric functions exist, we
are not aware of any example containing $\mbox{}_0F_{n-1}$.

\sect{Conclusions and outlook}\label{conc}

In this paper we have investigated the eigenvalue correlations for
the product of $n$ independent non-Hermitian random matrices.
Explicit results were given for any finite matrix size $N$
for all density correlation functions
in terms of 
the determinant of a kernel of orthogonal polynomials in the complex plane.
The corresponding weight functions were found to be
so-called Meijer $G$-functions depending on $n$,
whereas the joint probability distribution function remained the
absolute value squared of the Vandermonde determinant which is
standard for this symmetry class.

So far such results were only known for up to $n=2$ (in fact
for rectangular matrices with an elliptic law).
In the large-$N$ limit we investigated three different microscopic
limits where the local fluctuations were zoomed into: the 
microscopic edge, bulk and origin limit. 
In the former two cases we found a complete agreement with
the Ginibre ensemble of a single matrix $n=1$, after unfolding. The
edge and bulk limit are thus universal, as it was conjectured earlier
for the behaviour at the edge depending on the complementary error function.
In contrast at the origin each new matrix in the product adds a new
class of correlation functions, given by an $n$-dependent hypergeometric kernel
together with the Meijer $G$-function from the weight.

Several generalisations of the 
results presented here are
conceivable. First of all it should be possible to consider the product
of rectangular matrices instead. We expect that the results for the
edge and the bulk remain unchanged, whereas the origin limit will be
generalised as it is already known for $n=2$.
Matrices from other symmetry classes could be considered, such as
real quaternionic non-self dual or real asymmetric matrices. Again we
expect the edge and bulk behaviour to be unchanged, with local
microscopic modifications along the real and imaginary axis as well as
at the origin. Finally also for each matrix more general distributions 
than the Gaussian Ginibre distribution could be feasible. In
particular it would be interesting to know if the microscopic
properties we computed continue to hold 
for the product of non-Hermitian Wigner matrices with iid but 
non-Gaussian entries. Several of
these projects are currently under way.
\\

{\bf Acknowledgments:}
We would like to thank Eugene Strahov for useful discussions.
We acknowledge partial support by
the Polish Ministry of Science Grant No. N N202 229137 (2009-2012)
and by the Grant DEC-2011/02/A/ST1/00119 of the National Centre of Science
(Z.B.), as well as by the SFB $|$ TR12 ``Symmetries and Universality
in Mesoscopic Systems'' of the German research council DFG (G.A.).

\begin{appendix}

\sect{Computation of the Jacobian}\label{JacobianA}

We now calculate the Jacobi matrix for the change of variables from
$X$ to $(U,T,\Lambda)$. In order to
determine the relation between the infinitesimal elements (one-forms)
we differentiate both sides of (\ref{BLTU}). We obtain 
\begin{equation}
\begin{split}
dX & = \left(\begin{array}{ccc}
0 & dX_1 & 0 \\
0 & 0 & dX_2 \\
dX_3 & 0 & 0
\end{array}\right)
\\ & = \left(\begin{array}{ccc}
U_1 & 0 & 0 \\
0 & U_2 & 0 \\
0 & 0 & U_3
\end{array}\right)
\left(\begin{array}{ccc}
0 & dY_1 & 0 \\
0 & 0      & dY_2\\
dY_3 & 0 & 0 \\
\end{array}\right)
 \left(\begin{array}{ccc}
U^{-1}_1 & 0 & 0 \\
0 & U^{-1}_2 & 0 \\
0 & 0 & U^{-1}_3
\end{array}\right)\ ,
\end{split}
\label{dXdY}
\end{equation}
where
\begin{equation}
dY_j = d \Lambda_j +  d T_j + d M_j\ ,
\label{yltm}
\end{equation}
and
\begin{equation}
\begin{split}
\left(d M_1\right)_{ab} =  \left(dA_1\right)_{ab} x_{1b} - x_{1a}
\left( dA_2 \right)_{ab}
+ \sum_{c<b} \left(dA_1\right)_{ac} \left(T_1\right)_{cb} -
\sum_{d>a} \left(T_1\right)_{ad} \left(dA_{2} \right)_{db}\ ,\\
\left(d M_2\right)_{ab} =  \left(dA_2\right)_{ab} x_{2b} - x_{2a}
\left( dA_3 \right)_{ab}
+ \sum_{c<b} \left(dA_2\right)_{ac} \left(T_2\right)_{cb} -
\sum_{d>a} \left(T_2\right)_{ad} \left(dA_{3} \right)_{db}\ ,\\
\left(d M_3\right)_{ab} =  \left(dA_3\right)_{ab} x_{3b} - x_{3a}
\left( dA_1 \right)_{ab}
+ \sum_{c<b} \left(dA_3\right)_{ac} \left(T_3\right)_{cb} -
\sum_{d>a} \left(T_3\right)_{ad} \left(dA_{1} \right)_{db}\ .
\end{split}
\label{MA}
\end{equation}
The $dA_j = U^{-1}_j dU_j$ are anti-Hermitian matrices with
zeros on the diagonal. One can choose $\left(d A_j \right)_{ab}$ and
$\left(d A_j \right)^*_{ab}$ in the upper triangle ($a>b$)
to be independent infinitesimal elements of $dA_j$ (or more precisely -
independent one-forms). The elements $\left(d A_j \right)_{ab}$ in the
lower triangle ($a<b$), can be expressed by those in the upper one: 
$\left(dA_j\right)_{ab}=-\left(dA_j\right)^*_{ba}$.
The diagonal elements of $dA_j$ are zero since $U_j$
are restricted to $U(N)/U(1)^N$.

Now we are ready to calculate the Jacobian. Let us do it gradually.
First observe that the Jacobian for the change of variables
$dX = U dY U^{-1}$ eq. (\ref{dXdY}) equals one
since the matrix $U$ is unitary. So we have $|DX| = |DY|$.
Now we can calculate the Jacobian for the change from
$dY$ to ($d\Lambda, dT,dM$) in eq. (\ref{yltm}).
To this end let us write down (\ref{yltm}) in an explicit index notation
\begin{equation}
\begin{array}{llcll}
(dY_j)_{ab} & = d (\Lambda_j)_{ab} &                    &  + &
  (dM_j)_{ab}  \quad {\rm for} \ a=b \ ,\\ 
(dY_j)_{ab}  & =             & d (T_j)_{ab} & +  & (dM_j)_{ab}  \quad
  {\rm for} \ a<b \ ,\\ 
(dY_j)_{ab}  & =                     &    &  &  (dM_j)_{ab}  \quad
  {\rm for} \ a>b \ .
\end{array}
\label{YzTM}
\end{equation}
There are also corresponding equations for the complex conjugates.
The first line, for $a=b$, could be alternatively written as
$(dY_j)_{aa} = d x_{ja} + (dM_j)_{aa}$ since
$d(\Lambda_j)_{ab} = \delta_{ab} dx_{ja}$, or in shorthand notation
$d\Lambda=dx$. As one can see from (\ref{YzTM})
the Jacobi matrix for the linear transformation from the $dY$-basis to
the $(dx,dT,dM)$-basis is upper triangular and has 
a Jacobian equal one, so we have
\begin{equation}
|DX| = |DY| = |Dx| |DT| |DM| \ .
\end{equation}
The measure $DM$ in the last equation is a product of one-forms
$(dM_j)_{ab}$ for $a>b$.  They can be expressed as a
linear combination of the one-forms $dA$'s in eq. (\ref{MA}):
 \begin{equation}
|DX| = |Dx| |DT| |DA| \left| \frac{\partial M}{\partial A} \right|  \ .
\end{equation}
Let us calculate the Jacobian $|J| = \left| \partial M/\partial A \right|$
for this transformation (\ref{MA}). It has a specific
algebraic structure that imposes a certain ordering of the indices that
appear in (\ref{MA}). In fact this ordering implies that the
Jacobi matrix can be written as an upper triangular block matrix.
To see this let us first recall that we are considering only
the sector $a>b$ corresponding to the lowest line of (\ref{YzTM}).
Second of all, let us note that as a result of 
the $T_j$'s being strictly upper
triangular matrices we have further inequalities for the indices in  (\ref{MA}):
\begin{equation}
a>b>c  \quad  \mbox{and} \quad \ d>a>b \ .
\label{ineq}
\end{equation}
An inspection of the indices of the one-forms appearing in (\ref{MA}), 
$(dM_j)_{ab}$ ($a>b$), $(dA_j)_{ac}$ ($a>c$),  $(dA_j)_{db}$ ($d>b$), 
shows that they are all indexed by ordered pairs of indices whose
first index is larger than the second one. It is convenient to
introduce an increasing ordering in the set of such ordered pairs. A
pair $ab$ is said to be less than a pair $cd$ (and denoted by $ab<cd$) 
if $a>c$ or if $a=c$ and $b<d$. This choice of ordering
does not look very intuitive but it is convenient for our purposes.
For example for $N=4$ there are six ordered pairs and they are ordered
as follows $41 < 42 < 43 < 31 < 32 < 21$. We can now introduce a
single index $\alpha=1,\ldots,6$ that preserves the
increasing ordering of the pairs: $41 \rightarrow 1$, $42 \rightarrow
2$, $42 \rightarrow 3$, 
$31 \rightarrow 4$, $31 \rightarrow 5$, $21 \rightarrow 6$. Of course,
we can do this for any $N$. Using the index $\alpha$ we can concisely
write equations (\ref{MA}) as 
\begin{equation}
(d M_j)_\alpha = \sum_{j'\alpha'} (J_{jj'})_{\alpha\alpha'} (dA_k)_{\alpha'} \ .
\label{Jaa}
\end{equation}
Now we want to argue that the Jacobi matrix is upper block triangular,
that is $(J_{jj'})_{\alpha\alpha'} =0$ for all $\alpha>\alpha'$.
Clearly the matrix has diagonal blocks $\alpha=\alpha'$ (\ref{MA}).
It also has off-diagonal blocks coming from the sums on the right-hand side
of (\ref{MA}). We now show that the terms in the sums contribute to
the upper triangle $\alpha<\alpha'$. Indeed, as follows from the
inequalities (\ref{ineq})  the pairs of indices $ab$ of
$(dM_j)_{ab}$'s on the left hand side of (\ref{MA}) are smaller (in
the sense defined above: $ab<ac$ and $ab<db$) than the corresponding
pairs $ac$ and $db$ of $(dA_j)_{ac}$ and $(dA_j)_{db}$ in the sums 
on the right hand side of (\ref{MA}). This
is equivalent to saying that the sums run over $\alpha'$'s such that
$\alpha < \alpha'$. There are no terms for $\alpha > \alpha'$
on the right hand side of (\ref{MA}) and thus
the Jacobi matrix (\ref{Jaa}) $(J_{jj'})_{\alpha\alpha'}=0$ for
$\alpha>\alpha'$. Since the Jacobi matrix $(J_{jj'})_{\alpha\alpha'}$
is upper block triangular, its Jacobi determinant $|J|$ is identical
to the determinant of the corresponding block diagonal matrix
$(\widehat{J}_{jj'})_{\alpha\alpha'}$ that has the same diagonal
blocks: 
$|J| = |\widehat{J}|$. In this way we have reduced the calculation of
the determinant $|\partial M/\partial A|$
to the calculation of the determinant $|\partial \widehat{M}/\partial A|$
of the block diagonal matrix $\widehat{J}$ corresponding to the
transformation obtained from (\ref{MA}) by skipping upper-triangular
terms, that is the sums: 
\begin{equation}
\begin{split}
\left(d \widehat{M}_1\right)_{ab} & =  \left(dA_1\right)_{ab} x_{1b} - x_{1a}
\left( dA_2 \right)_{ab}\ , \\
\left(d \widehat{M}_2\right)_{ab} & =  \left(dA_2\right)_{ab} x_{2b} - x_{2a}
\left( dA_3 \right)_{ab}\ ,  \\
\left(d \widehat{M}_3\right)_{ab} & =  \left(dA_3\right)_{ab} x_{3b} - x_{3a}
\left( dA_1 \right)_{ab}  \  .
\end{split}
\label{diag1}
\end{equation}
The Jacobian can be calculated by the use of the external (wedge) product
of one-forms $(d\widehat{M}_j)_{ab}$. 
Let us first calculate it for a given pair $ab$. 
Using (\ref{MA}) we have:
\begin{equation}
\begin{split}
& (d\widehat{M}_1)_{ab} \wedge (d\widehat{M}_2)_{ab} \wedge 
(d\widehat{M}_3)_{ab} = \\
& (dA_1)_{ab} \wedge (dA_2)_{ab} \wedge (dA_3)_{ab}
\big(x_{1b} x_{2b} x_{3b} - x_{1a} x_{2a} x_{3a}\big) \ .
\end{split}
\label{diag3}
\end{equation}
Taking also into account the corresponding independent equations
for the complex conjugates, we obtain a transformation of the $6$-form
in the whole $ab$ sector~:
\begin{equation}
\begin{split}
& (d\widehat{M}_1)_{ab} \wedge (d\widehat{M}_2)_{ab} \wedge
  (d\widehat{M}_3)_{ab} \wedge 
(d\widehat{M}_3)^*_{ab} \wedge (d\widehat{M}_2)^*_{ab} \wedge
  (d\widehat{M}_1)^*_{ab} = 
\\
& (dA_1)_{ab} \wedge (dA_2)_{ab} \wedge (dA_3)_{ab} \wedge
(dA_3)^*_{ab} \wedge (dA_2)^*_{ab} \wedge (dA_1)^*_{ab} 
\\
& \times\big|x_{1b} x_{2b} x_{3b} - x_{1a} x_{2a} x_{3a}\big|^2 \ .
\end{split}
\end{equation}
We repeat this calculation independently for all sectors $a>b$ and
eventually obtain the volume
form:
\begin{equation}
|DM| = |J(x)| |DA| = \prod_{a>b}^N
\big|x_{1b} x_{2b} x_{3b} - x_{1a} x_{2a} x_{3a}\big|^2 |DA| \ ,
\end{equation}
and
\begin{equation}
|DX| =  |J(x)| |Dx| |DT| |DA| \ .
\end{equation}

We are ready to write down the partition function (\ref{Z}) in new variables, for general $n$.
The Gaussian weight function (\ref{Z}) assumes a form:
\begin{equation}
\e^{- {\rm Tr} \sum_{j=1}^n X^{\dagger}_j X_j} =
\e^{- \sum_{j=1}^n\sum_{a=1}^N |x_{ja}|^2} \e^{-
  \sum_{j=1}^n\sum_{a>b}^N \left|(T_j)_{ab}\right|^2} \ ,
\end{equation}
that is independent of $dA_j$'s and thus of $dU_j$'s:
$dA_j = U_j^{-1} dU_j$. So we can integrate out the $dU$-variables. This
integration gives a constant 
equal to the volume of the coset $U(N)/U(1)^N$. Also the integration
over the $(T_j)_{ab}$'s can be done since it is an independent Gaussian
integral. By comparing to Ginibre we obtain for each $j$
\be
\int |DU_j|\int|DT_j|\exp[-\Tr\ T^\dag_j T_j]=1/(N!\ G[N+1]\pi^N)=C_1.
\ee
Denoting by $C_n$ the total constant factor coming
from the integration over all $dU_j$'s and $dT_j$'s, we have $C_n=(C_1)^n$.
What remains is an integral over the $dz$'s as was claimed in section \ref{jpd}:
\begin{equation}
Z_n=C_n \int \prod_{j=1}^n\prod_{a=1}^N d^2x_{ja} \prod_{a>b}^N
\big|x_{1b} x_{2b} x_{3b} - x_{1a} x_{2a} x_{3a}\big|^2\ 
\e^{- \sum_{j=1}^n\sum_{a=1}^N |x_{ja}|^2}\ .
\end{equation}

\sect{Asymptotic of the weight function}
\label{weightB}

In order to make our paper self contained we compute here
the leading order asymptotic behaviour of
the weights $w_n(z)$ for $|z| \rightarrow \infty$ using the saddle
point method. The result can also
be found in Theorem 2 in reference \cite{Fields}.

We use the multidimensional representation of the
weights (\ref{wn-def}):
\begin{equation}
w_n(z) = (2\pi)^{n-1} \int \prod_{j=1}^{n-1} (dr_j/r_j)
\ \e^{-S}\ ,\ \ n>1\ ,
\end{equation}
where
\begin{equation}
S = \frac{|z|^2}{(r_1\ldots r_{n-1})^2} + \sum_{j=1}^{n-1} r_j^2 \ .
\end{equation}
The saddle point equation
\begin{equation}
\frac{\partial S}{\partial r_j} = \frac{-2}{r_j}
\frac{|z|^2}{(r_1\ldots r_{n-1})^2} + 2 r_j = 0
\end{equation}
has a symmetric solution $r_1=r_2 = \ldots r_{n-1} \equiv r_*=|z|^{1/n}$.
The equation has no other solutions, so the symmetric solution
is not only a local minimum of $S$ but also the global one.
At the minimum the function $S$ assumes the value $S_* = n |z|^{2/n}$.
The Hessian
$H_{ij} = \frac{\partial^2 S}{\partial r_i \partial  r_j}
=\frac{(4+2\delta_{ij})|z|^2}{r_ir_j(r_1\ldots r_{n-1})^2}+2\delta_{ij}$
takes the following values at the minimum $r_*$:
$H_{*ij}= 4$ for $i\ne j$ and $H_{*ij} = 8$ for $i=j$. The
determinant of the Hessian is $\det [H_*] = n\, 4^{n-1}$. So we can write
an explicit formula
for the leading order asymptotic behaviour of $w_n(z)$ for large $z$ as
\begin{equation}
w_n(z) \sim \left(\frac{2\pi}{r_*}\right)^{n-1}
\frac{(2\pi)^{(n-1)/2}}{(\det H_*)^{1/2}} \e^{-S_*}
\end{equation}
that gives
\begin{equation}
w_n(z) \sim \frac{1}{\sqrt{n}}
\left(2\pi^3\right)^{(n-1)/2} |z|^{(1-n)/n} \e^{-n |z|^{2/n}} \ ,
\label{asympt}
\end{equation}
which agrees with \cite{Fields} where also the error terms are computed.
In particular for $n=2,3$ we have
\begin{equation}
w_2(z) \sim \pi^{3/2} z^{-1/2} \e^{-2|z|}
\ , \ w_3(z) \sim \frac{2\pi^3}{\sqrt{3}} |z|^{-2/3}
\e^{-3|z|^{2/3}}\ ,
\end{equation}
where the first result $n=2$ agrees with the known asymptotic of
$K_0(2|z|)$ from eq. (\ref{w2-Osb}) \cite{Grad}.
In fact eq. (\ref{asympt}) also holds for $n=1$ where $w_n(z) \sim
\exp[-|z|^2]$ is exact.

The asymptotic formula (\ref{asympt}) is consistent with the
recurrence relation (\ref{wn-rec}) in the sense that when one inserts
the asymptotic form of $w_n(z)$ eq. (\ref{asympt}) into the integral on the
right hand side of the recurrence relation (\ref{wn-rec}) one obtains
the asymptotic form of $w_{n+1}(z)$ as given in (\ref{asympt}).

\sect{Asymptotic of the hypergeometric kernel}\label{kernelC}

In this appendix we discuss the asymptotic behaviour
of the truncated 
sum in the kernel eq. (\ref{Ker-result}).
The asymptotic of the corresponding infinite sum was discussed 
in reference \cite{BJK}, and
we again give some details here to be self-contained. Let us repeat
the definition
\begin{equation}
T_n(x,N) = \sum_{k=0}^{N-1} \frac{x^k}{(\pi k!)^n}\ ,
\label{tsum}
\end{equation}
which we would like to investigate
for large $N$ and for $x$ of order $N^n$, due to $x=|z|^2\lesssim N^n$.
We again apply the saddle point method. By first
using Stirling's formula $k!\approx \sqrt{2\pi k} (k/e)^k$ and then
approximating the sum by an integral we obtain:
\begin{equation}
T_n(x,N) \approx (2\pi^3)^{-n/2} \int_1^N dk  k^{-n/2} \e^{s(k)}\ ,
\end{equation}
where 
$s(k)=k \ln[x] - n k \ln [k] + nk$. We are interested in the
behaviour 
of $T_n(x,N)$ for large $N$ and $x$. In this case the exponent $\e^{s(k)}$
can be approximated by a Gaussian function with a maximum
located at
\begin{equation}
k_* = x^{1/n}\ ,
\label{max}
\end{equation}
being a solution of the saddle point equation
\begin{equation}
s'(k_*) = \ln [x] - n \ln [k_*] = 0 \ .
\end{equation}
We also have $s(k_*)=nk_*$, $s''(k_*) = -n/k_*$, so the
Gaussian approximation gives:
\begin{equation}
T_n(x,N) \approx (2\pi^3)^{-n/2} \e^{nk_*}
\int_1^N dk  k^{-n/2} \e^{-n(k-k_*)^2/(2k_*)}\ ,
\label{ti}
\end{equation}
with $k_*=x^{1/n}$ (\ref{max}). Let us change the integration
variable in the last integral to $t= \sqrt{n/2k_*}\ (k-k_*)$.
Written in this new variable the Gaussian part assumes 
the form $\e^{-t^2}$,
which means that values of $t$ that contribute to the integral are of
order one.
In this narrow range of $t$, the factor $k^{-n/2}$ can be treated as
constant. Indeed, for $t$ of order of unity and $k_*$ of order $N$, we
have
$k = k_*(1 + t/(nk_*/2)^{1/2} \approx k_*$, for $t\ll (nk_*/2)^{1/2}$.
The upper integration limit for $t$ is: $\sqrt{n/2k_*}(N-k_*)$. The lower one
is far below the range of the Gaussian integrand, so it can be set to
$-\infty$.  Replacing $k_*$ by $x^{1/n}$ (\ref{max}) we eventually obtain
\begin{equation}
\lim_{N>k_*\gg1}
T_n(x,N) \approx \frac{1}{\pi\sqrt{n}} \left(2\pi^3\right)^{(1-n)/2}
x^{(1-n)/2n} \exp\left[nx^{1/n}\right]
 \frac{1}{2} \mbox{erfc}\left(
\frac{\sqrt{n}\left(x^{1/n}-N\right)}{\sqrt{2x^{1/n}}}\right) \ .
\label{T-asympt}
\end{equation}
In the limit $N\rightarrow \infty$ the truncated sum becomes
an infinite series $T_n(x) = \lim_{N\rightarrow \infty} T_n(x,N)$.
In this case the saddle point is always located within the range
of integration, so we have
\begin{equation}
T_n(x) = \sum_{k=0}^{\infty} \frac{x^k}{(\pi k!)^n}
\stackrel{x\gg1}{\approx}
\frac{1}{\pi\sqrt{n}} \left(2\pi^3\right)^{(1-n)/2}
x^{(1-n)/2n} \exp[nx^{1/n}] \ .
\label{TNinfinity}
\end{equation}

\sect{Eigenvalue distribution of n powers of a single Ginibre matrix}
\label{nGinibreD}

In this appendix we discuss the eigenvalue distribution of the $n$-th power
of a single Ginibre matrix. Consider a random matrix $A$ with a
spherically symmetric eigenvalue distribution $\rho(w) = \rho(|w|=r)$.
Denote by $\tilde{\rho}_n(w)$ the corresponding spectral density of $A^n$ -
the $n$-th power $A$, which is also angle-independent.
If $\lambda$ is an eigenvalue
of $A$, then $\tilde{\lambda}=\lambda^n$ is an eigenvalue of $A^n$.
Changing the integration variable in the formula below to $s=r^n$,
we have:
\be
\int_0^\infty 2\pi r \rho(r) dr = \int_0^\infty \frac{2\pi
  s^{2/n-1}}{n} \rho\left(s^{1/n}\right) ds = \int_0^\infty 2\pi s
\tilde{\rho}_n\left(s\right) ds \ ,
\ee
and thus
\be
\tilde{\rho}_n(r) = \frac{r^{\frac{2}{n}-2}}{n} \rho\left(r^{1/n}\right) \ .
\label{power}
\ee
In particular, for the limiting macroscopic density of
a single Ginibre matrix $X$ with $n=1$
we have $\rho_{macro}^{(1)}(w) = \frac{1}{\pi} \Theta(1-|w|)$ and
therefore the corresponding density of its $n$-power is
\be
\tilde{\rho}_{macro}(w) = \frac{|w|^{\frac{2}{n}-2}}{\pi n} \Theta(1-|w|)\ .
\ee
This is
exactly as for the product of $n$-independent matrices (\ref{limiting-n}).
We can also repeat calculations for the finite size distribution
for a single Ginibre matrix in eq. (\ref{fs-n}) :
\be
\rho(w)=N^{-1}R_1^{(1)}(N^{\frac12}w) =
\frac{1}{2\pi} \mbox{erfc}\left(\sqrt{2N}\left(|w| - 1\right)\right) \ .
\ee
Applying (\ref{power}) we find the corresponding distribution
of eigenvalues of its $n$-th power:
\be
\tilde{\rho}_{n}(w)
=  \frac{|w|^{\frac{2}{n}-2}}{\pi n}  \frac{1}{2}\mbox{erfc}
\left(\sqrt{2N}\left(|w|^{1/n} - 1\right)\right) \ .
\ee
It can be further simplified using a Taylor expansion at $|w|=r=1$ to
approximate the function inside the erfc-function: $r^{1/n}-1 =
(1/n)(r-1) + \ldots$ and neglecting higher order terms, as in section
\ref{rho_macro}:
\be
\tilde{\rho}_{n}(w)
=  \frac{|w|^{\frac{2}{n}-2}}{\pi
  n}  \frac{1}{2}\mbox{erfc}
\left(\frac{\sqrt{2N}}{n}\left(|w| - 1\right)\right) \ .
\ee
An important difference between this case and the power of $n$ independent
matrices is that the dependence on $n$ in the denominator is
$n$, and not $\sqrt{n}$ as in eq. (\ref{fs-n}).

\end{appendix}


\end{document}